\documentclass[letterpaper,12pt]{revtex4}
\usepackage{amsfonts}
\usepackage{amsmath}
\usepackage{amssymb}
\usepackage{graphicx}

\begin{document}
\title{Interference of surface plasmon polaritons from a "point" source}
\author{Xifeng Ren}
\email{renxf@ustc.edu.cn}
\author{Aiping Liu}
\author{Changling Zou}
\author{Lulu Wang}
\author{Yongjing Cai}
\author{Fangwen Sun}
\author{Guangcan Guo}
\author{Guoping Guo}
\affiliation{Key Laboratory of Quantum Information, University of
Science and Technology of China, Hefei 230026, People's Republic of
China}


\begin{abstract}
  The interference patterns of the surface plasmon polaritons(SPPs) on the metal surface from a "point" source are observed.
  These interference patterns come from the forward SPPs and the reflected one from the obstacles, such as straightedge,
 corner, and ring groove structure. Innovation to the previous works, a "point" SPPs source with diameter
  of $100$ nm is generated at the freely chosen positions on Au/air interface using near field excitation method.
  Such a "point" source provides good enough coherence to generate obvious interference phenomenon. The constructive and destructive
  interference patterns of the SPPs agree well with the numerical caculation. This "point" SPPs source may be useful in the investigation
  of plasmonics for its high coherence, deterministic position and minimum requirement for the initial light source.
\end{abstract}
\maketitle
\section{Introduction}
Surface plasmon polaritons (SPPs), collective oscillating electrons
excited by electromagnetic field, have been studied for decades and
a great of interest has been injected into this area. Such SPPs are
involved in a wide range of phenomena \cite{Barnes03,Ozbay06},
including nanoscale optical waveguiding
\cite{Bozhevolnyi06,Takahara97,Takahara04,Zia06}, perfect lensing
\cite{Pendry00}, extraordinary optical transmission
\cite{Ebbesen98}, subwavelength lithography \cite{Fang05}, and
ultrahigh sensitivity biosensing \cite{Liedberg83}. It has also been
experimentally proved that SPPs are useful in the investigation of
quantum information \cite{Altewischer02,Fasel05,Ren06}. To realize
the full potential technology of plasmonics, we need to construct a
general framework to describe the propagation, diffraction, and
interference of SPPs. The interference phenomenon of SPPs has been
studied in many works \cite{Lezec06,Lopez07,Paci07,Zia07,Ren08}, for
example, a double-slit experiment with SPPs is presented, which
reveals the analog between SPPs propagating along the surface of
metallic structures and light propagating in conventional dielectric
components \cite{Zia07}.

In the investigation of plasmonics, a good source of SPPs is very
important which proposes high demands on the exciting light source.
Usually the SPPs are excited by far-field excitation using grating
coupling or attenuated total reflection (ATR). The spots of the
focused light always have the diameter bigger than half of the light
wavelength due to diffraction limit. As we know, one of the most
important properties of light source is the coherence, which is
essential in the study of the interference. For a non-point source,
the spatial coherence is correlated with the size of source, such as
the case in the double-slit interference experiment. Another
standing problem for the SPPs source is the precisely generation of
SPPs at designated positions of
nanostructures\cite{Dall09,Bouhelier07}, which is very important in
usage of plasmonic devices. In this paper, a "point" SPPs source is
generated by a near-field scanning optical microscope (NSOM) probe
with a diameter of few hundred nanometers, which is smaller than the
wavelength of SPPs. With the benefit of the near field scanning
method, the source can also be placed at any positions of the
nanostructure precisely. The interference patterns between the
forward SPPs and the reflected one from the obstacles, such as
straightedge, corner, and ring groove structure are detected by
another NSOM probe connected to an avalanche photo diode (APD). The
"point" SPPs source makes the interference pattern clear due to its
small size. This new method will benefit the study of the plasmoincs
where a SPPs source with high coherence and deterministic position
is needed. Since the SPPs only propagate on the surface, the same as
the wave on the water surface, our results may also be useful in the
investigation of some fundamental physics by simplifying the model
from three dimensional space to two dimensional space.

\section{Results and discussion}

In this literature, the intensity distribution of SPPs is studied by
the near-field optical measurements with $MultiView 4000^{TW}$
 scanning dual probe microscope/NSOM system (Nanonics Imaging Ltd.,
Jerusalem, Israel). This system allows for simultaneous atomic force
microscopy (AFM) and NSOM imaging. A sketch map of our experimental
setup is given in Figure 1. The SPPs is generated by a cantilevered
aperture NSOM probe (diameter 100 nm or 500 nm), which is connected
with a semiconductor laser with 670 nm wavelength using a single
mode polarization maintaining fiber. The NSOM probe is coated with
Cr-Au(0.02 and $0.2 \mu m$) to prevent light emitting from the side
of the fiber cone and concentrate light to emit from the aperture.
When the probe approaches the metal surface with the distance of
only a few nanometers (called near field), the SPPs can be excited
effectively\cite{Dall09}. This makes the light source small enough
to be a "point" source which avoids the shortcomings that light
diffract from the fiber cone with a big scatting spot
\cite{Bouhelier07}. A similar 100 nm width aperture NSOM probe scan
the sample in collection mode to measure the field intensity
associated with the interference pattern of the propagating SPPs
waves. It is finally collected by a single photon detector (SPCM-15,
PerkinElmer Optoelectronics, Canada). The metal plate used in our
experiment to excite SPPs is produced as follows. After subsequently
evaporating a 3-nm titanium bonding layer, a $1 \mu m$ thick gold
layer is evaporated onto a 0.5-mm-thick silica glass substrate. An
initial AFM scan of the sample was performed, which gave a root mean
squared roughness of 2 nm.  The gold layer is thick enough to avoid
the excitation of SPPs on the interface of $Au/SiO_{2}$.
\begin{figure}
\includegraphics[width=4in]{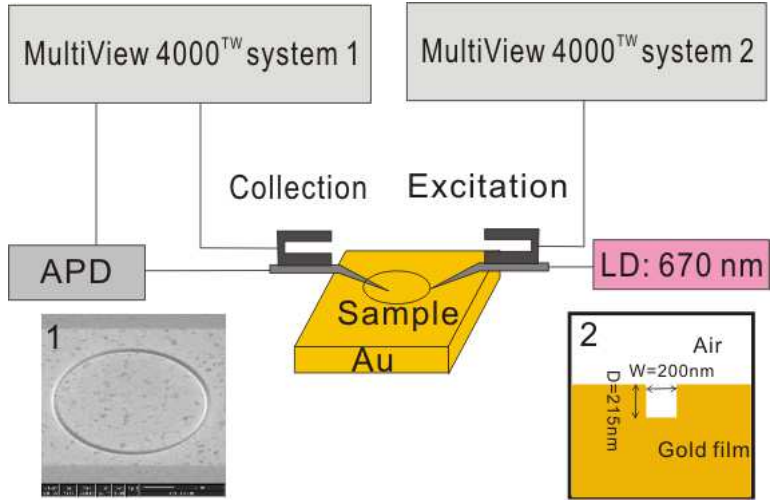}\newline\caption{(Color online)
Schematic setup of the experiment for the near-field excitation and
near-field collection of SPPs wave at Au/air interface. The inset 1
is the SEM picture of the ring-shaped groove and the inset 2 gives
the cross section of the ring-shaped groove.}
\end{figure}

Firstly, we discuss the relation between the visibility of
interference patterns and the size of the "point" source. According
to the coherence theory, the better the coherence condition is met,
the higher visibility of the interference fringes can be achieved.
We test this by observing the SPPs interference phenomenon of a ring
structure with different "point" sources size. The ring-shaped
groove structure is produced with focused ion beam etching system
(DB235 of FEB Co.), which has a groove depth of 215 nm, width of 200
nm and diameter of $10 \mu m$(see inset of Figure 1). The SPPs is
excited by the laser light of 670 nm wavelength through the
excitation NSOM probe and another NSOM probe with diameter of $100$
nm works on the collection mode to scan the sample and image the
SPPs distribution on the metal surface. When the excitation probe
approaches the metal plate, the excited SPPs propagate along the
interface of the Au/air and strike the ring edge. According to the
Huygens' principle, every point of the groove's edge can be regarded
as a new "point" SPPs source, from which the SPPs form the
interference pattern in the observed area. Two excitation NSOM
probes with aperture diameters 100 nm and 500 nm are used and the
NSOM images are given in Figure 2a and 2b respectively. The
interference pattern in Figure 2a is much clearer than that in
Figure 2b which confirms that a NSOM probe with smaller aperture
produces a higher coherent SPPs source. Figure 2c gives the SPPs
intensity distribution along the blue line in Figure 2a and 2b,
which also shows an obvious difference between the two cases. For a
further study, we also change the diameter of the ring to $5 \mu m$
with the same excitation probe of $100$nm diameter. The NSOM image
of the SPPs distribution is given in Figure 2d which possesses the
same interference pattern but less clear. Since the probe with $100
nm$ diameter can act as a better "point" SPPs source, we use it to
observe the interference patterns of SPPs in the following
experiments.

\begin{figure}
\includegraphics[width=4in]{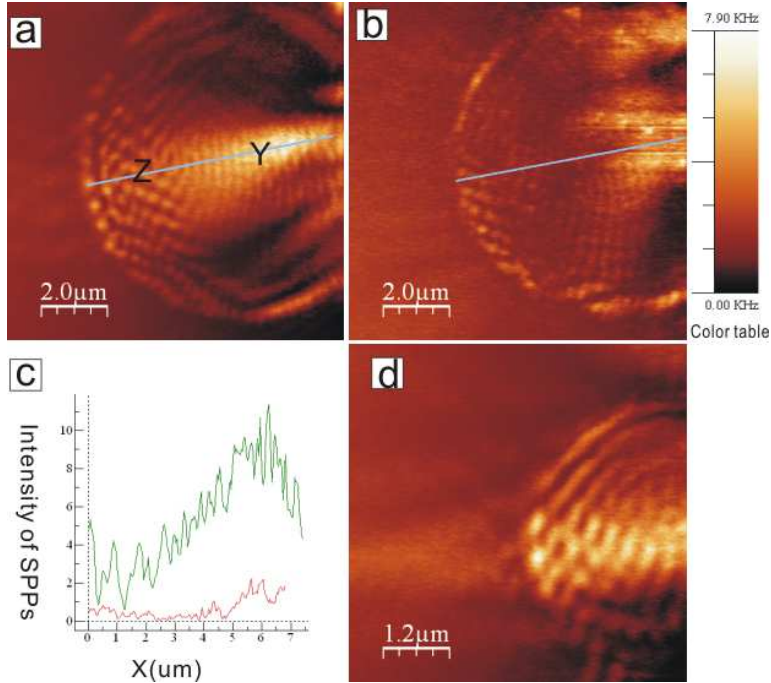}\newline\caption{(Color online)
NSOM images of the SPPs distribution on the ring-shaped groove
structure with ring diameter $r$ and the excitation probe diameter
$d$. (a) $r=10 \mu m$, $d=100 nm$; (b) $r=10 \mu m$, $d=500 nm$; (d)
$r=5 \mu m$, $d=100 nm$. (c) The green line and red line correspond
to the intensity distribution of SPPs along the lines in Figure 2a
and 2b respectively.}
\end{figure}
Besides the small size of the SPPs source, another benefit of our
method is the deterministic source position. We can move the
excitation probe to any positions of the sample by performing an
initial AFM scan. This is illustrated by observing the interference
phenomenon of the ring structure with different SPPs source
positions, as shown in Figure 3a. The SPPs is excited at the
positions of "b", "c", "e" and "g" along the red line, where "b" is
the center of the ring, "c" and g are $2 \mu m$ inner and out of the
ring structure and "e" is inside the ring groove. Another NSOM probe
works on the collection mode to scan the area marked by dished
square and image the SPPs distribution on the sample surface. The
interference pattern changes with different excitation positions as
shown in Figure 3b, c, e and g accordingly. A simple description is
given as follows. When the excitation probe is put at the center of
the ring structure (position "b"), the distance between the point
source and the ring structure is almost the same. A ring shaped
interference patterns are observed as shown in Figure 3b. The
inhomogeneity may come from the imperfect NSOM probe. This is
verified by scanning a flat metal surface near the excitation probe
(see the supporting information), which also give a similar
asymmetric energy distribution. This ring shaped interference
patterns change to curves when SPPs are generated at "c". When the
excitation probe move to position "g", only the right part of the
groove acts as new SPPs sources. The SPPs generated at the groove
edge propagate from right to left as shown in Figure 3g. At the
right part inside the ring, the fringe is curved since the wall of
the groove is curved. While at the left part, especially nearby the
groove, the interference pattern looks like grid, which is formed by
the interference of the forward SPPs and the reflected ones from the
left groove. More interestingly, the image is much clear when the
excitation probe is put inside the groove as shown in Figure 3e. At
this time, guiding SPPs are excited in the groove and transmit along
the groove waveguide\cite{Bozhevolnyi06}. The leakage parts from the
groove edge act as new sources. Obviously, the intensities of these
new sources are much higher than the cases that excitation probe is
outside the groove. This gives a brighter interference pattern.
Besides the interference patters of the SPPs, an intensity peak
appears at the center of the ring, because of the SPPs focusing
effect of the ring-shaped groove\cite{Liu05}. Numerical calculations
are also given for the cases that excitation probe is put the
positions "c","e" and "g", as shown in Figure 3d, f and h. They are
similar with the experimental results.
\begin{figure}
\includegraphics[width=3in]{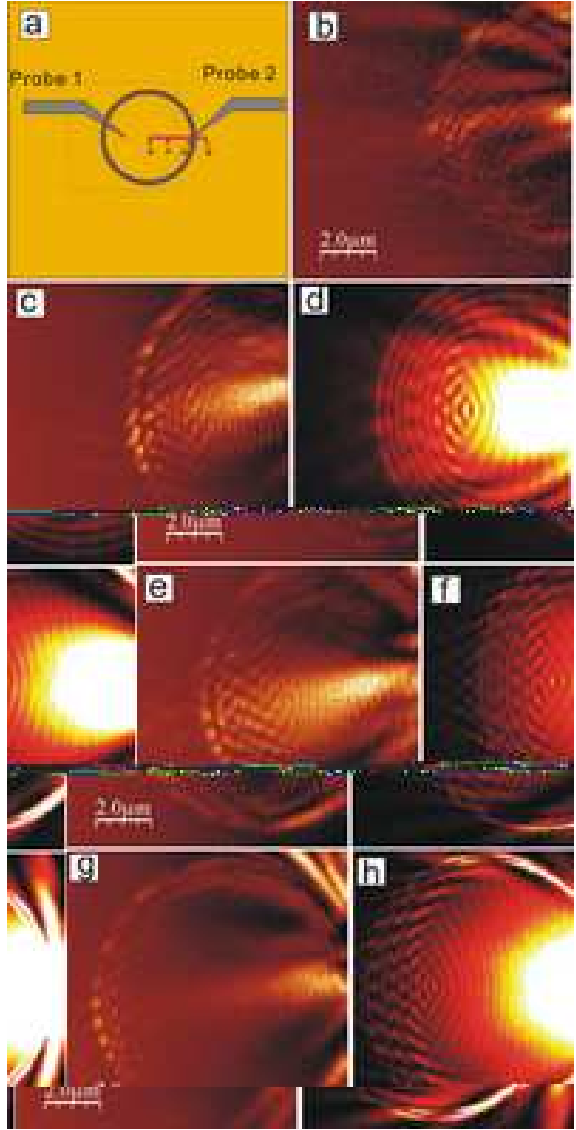}\newline\caption{(Color online)
(a) The schematic map of the ring-shaped groove and different
excitation positions. (b),(c),(e) and (g) NSOM images of the SPPs
distribution on the sample with excitation position at "b", "c", "e"
and "g" along the red line shown in Figure 3a. (d),(f) and (h)
Numerical calculations of the energy distribution corresponding to
Figure 3(c),(e) and (g) respectively.}
\end{figure}

To give a more clear picture of this interference phenomenon, we
also use more simple metal structures, for example, a straight edge
and a corner. We cut two cross slots on the gold film with the width
of $30 \mu m$, which is big enough to stop the SPPs coupling between
two edges of the slot. The sample is excited at point S, as shown in
inset of Figure 4a. Another NSOM probe (collection probe) scans the
area of $5 \mu m
*5 \mu m$ marked by "b". The positions of the two probes and the
distance between them can be controlled freely by the two SPM
controllers of the $MultiView 4000^{TM}$ system. The SNOM image is
shown in Figure 4b and the fringe is formed by the interference of
the propagating SPPs and the reflected one from the vertical edge of
the gold film, since the SPPs reflected from the farther horizontal
edge can be neglected at this area. Figure 4d is the SPPs intensity
distribution along the blue line in Figure 4b, which shows that the
average period of the stripes is about 320 nm. This is half of the
wavelength of the SPPs, which is 640 nm calculated from the function
of
\begin{equation}
    \lambda_{spp}=\lambda_{0}[(\varepsilon_{d}+\varepsilon_{m}) /(\varepsilon_{d} \varepsilon_{m} ) ]^{1/2}
\end{equation}
where $\lambda_{spp}$ and $\lambda_{0}$ are the wavelength of the
SPPs and the light in vacuum, $\varepsilon_{d}$ and
$\varepsilon_{m}$ are the real parts of the metal and dielectric
permittivities, respectively. Next, the collection probe 2 scans the
area of $6 \mu m *6 \mu m$ marked by "c" with the excitation probe
at position S as former. The SNOM image is shown in Figure 4c, in
which we can see grids obviously. In this case, the distances
between the scanned area and the two edges are about the same, so
the SPPs waves reflected from the horizontal edge must also be
considered. To give a simple description, we establish a
two-dimensional coordinate system with the excitation position S as
the origin on the surface of Au film as shown in Figure 4a. At an
arbitrary point $A(x,y)$, if we only consider the impact of the edge
along the $X$ direction, the interference comes from the two SPPs
waves from the point S and its symmetric point $S_{1}$ with the
vertical edge. Obviously, the interference fringes should be
parallel to the edge approximately and the period is given by
$\lambda_{spp}/2$ . Now, plus the impact of another adjacent edge,
there are four symmetric point sources with the two cross edges, S,
$S_{1}$, $S_{2}$ and $S_{3}$ as shown in Figure 4a. The total
electric field at point A is
\begin{equation}
    E_{all}=E_{0}(x,y)+E_{1}(x,y)+E_{2}(x,y)+E_{3}(x,y)
\end{equation}
where $E_{0}(x,y)$, $E_{1}(x,y)$,$E_{2}(x,y)$ and $E_{3}(x,y)$ are
the electric field of SPPs waves at point A from "point" sources of
S, $S_{1}$, $S_{2}$ and $S_{3}$ respectively. This interference
pattern is more complex than the case only considering the
superposition of two waves. But in principle, the interference
pattern should be grid in the area where the SPPs source from four
sources have to be considered simultaneously.
\begin{figure}
\includegraphics[width=4in]{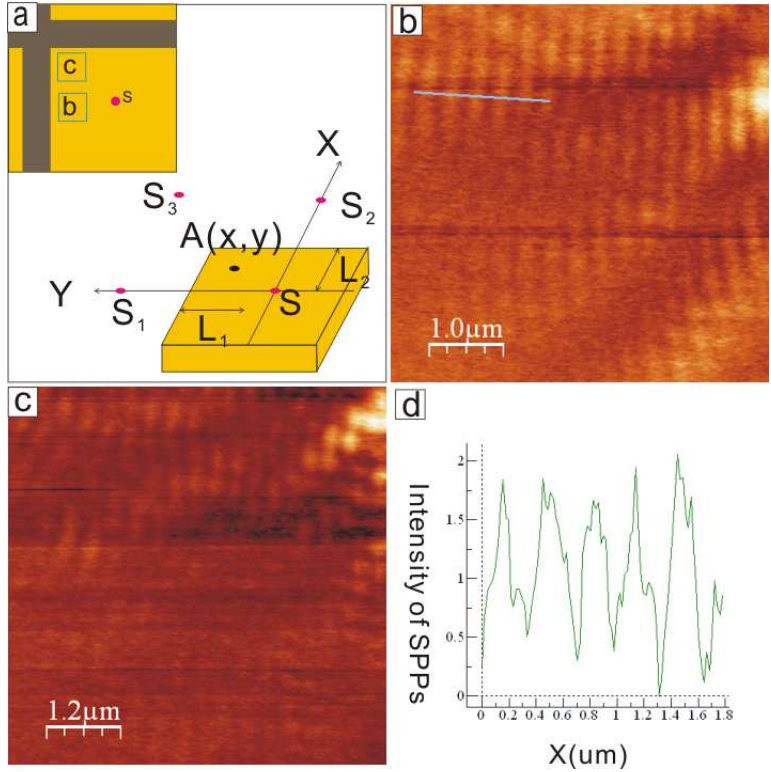}\newline\caption{(Color online)
(a) A two-dimensional coordinate system with S as the origin on the
surface of Au film, the inset is a schematic of two cross slots on
the Au film surface. (b)and (c) NSOM images of the SPPs distribution
on the area marked by the square "b" and "c" respectively as shown
in the inset of Figure 4a. (d) The intensity distribution along the
line in Figure 4b gives a interference period of about 320 nm.}
\end{figure}

\section{Conclusion}

In this literature, a "point" source is used to study the
interference of the SPPs. The interference patterns of strip, grid
and ring are observed on the metal surface. The interference pattern
appears more clear when the coherence condition is met better by
using a smaller SPPs source size. The success of the experiment is
benefited from the use of the NSOM probe to excite the sample in the
near field. This "point" SPPs source can be used in many areas for
its high coherence, deterministic position and minimum requirement
for the initial light source.

\subsection{Methods}
\textbf{Preparation of Samples.} The metal plate is prepared by
subsequently evaporating a 3-nm titanium bonding layer and a $1 \mu
m$ gold layer onto a 0.5-mm-thick silica glass substrate. We cut two
cross slots on the gold film with a width of about $30 \mu m$ by the
tip of a knife and then the corner and edges are formed. The ring
shaped groove with depth of 215 nm, width of 200 nm and diameter of
$10 \mu m$ was produced with focused ion beam etching system (DB235
of FEB Co.).

\textbf{SPPs Excitation and Detection.} A semiconductor laser of 670
nm wavelength is connected to the cantilevered NSOM probe by a
single mode polarization maintaining fiber. The polarization of the
laser is controlled by a polarizer. The metal sample of two cross
slots is excited at the point S by the excitation NSOM probe. The
other similar NSOM probe with a 100 nm width aperture scans the
areas of $5 \mu m*5 \mu m$ marked by "b" and $6 \mu m *6 \mu m$ by
"c" respectively in collection mode and measures the field intensity
associated with the interference pattern of the propagating SPPs
waves. The position of the two probes and the distance between thems
are controlled by the two SPM controllers of the $MultiView
4000^{TM}$ system. The collection NSOM probe is connected to a
single photon detector (SPCM-15, PerkinElmer Optoelectronics,
Canada) through another single mode polarization maintaining fiber.
For case of the ring groove structure with a diameter of $10 \mu m$,
the experimental setup is the same as the former. The excited points
change along the red line in Figure 3a and the collection probe
scans the left part of the ring each time. We also scan a ring
groove structure with a diameter of $5 \mu m$ using the same method.

This work was funded by the National Basic Research Programme of
China (Grants No.2011CBA00200), the Innovation funds from Chinese
Academy of Sciences, and the National Natural Science Foundation of
China (Grants No.10874163 and No.10904137),Anhui Provincial Natural
Science Foundation(Grants No. 090412053) and Science and
Technological Fund of Anhui Province for Outstanding Youth(Grants
No.2009SQRZ001ZD).

Figure S1 show the NSOM image of the SPPs distribution on the gold
plane near the point SPPs source without obstacles; Figure S2 show
the NSOM images of the SPPs distribution on the the ring groove
structure with a diameter of $5 \mu m$ excited by the 670 nm and 633
nm wavelength laser respectively.


\begin{thebibliography}{99}
\bibitem {Barnes03}Barnes, W. L.; Dereux, A.; Ebbesen, T. W., Surface plasmon subwavelength optics. Nature 2003, 424 (6950), 824-830.
\bibitem {Ozbay06}Ozbay, E., Plasmonics: Merging Photonics and Electronics at Nanoscale Dimensions. Science 2006, 311 (5758), 189-193.
\bibitem {Bozhevolnyi06}Bozhevolnyi, S. I.; Volkov, V. S.; Devaux, E.; Laluet, J.-Y.; Ebbesen, T. W., Channel plasmon subwavelength waveguide components including interferometers and ring resonators. Nature 2006, 440 (7083), 508-511.
\bibitem {Takahara97}Takahara, J.; Yamagishi, S.; Taki, H.; Morimoto, A.; Kobayashi, T., Guiding of a one-dimensional optical beam with nanometer diameter. Opt. Lett. 1997, 22 (7), 475-477.
\bibitem {Takahara04}Takahara, J.; Kobayashi, T., Low-Dimensional Optical Waves And Nano-Optical Circuits. Opt. Photon. News 2004, 15 (10), 54-59.
\bibitem {Zia06}Zia, R.; Schuller, J. A.; Chandran, A.; Brongersma, M. L., Plasmonics: the next chip-scale technology. Materials Today 2006, 9 (7-8), 20-27.
\bibitem {Pendry00}Pendry, J. B., Negative Refraction Makes a Perfect Lens. Phys. Rev. Lett. 2000, 85 (18), 3966.
\bibitem {Ebbesen98}Ebbesen, T. W.; Lezec, H. J.; Ghaemi, H. F.; Thio, T.; Wolff, P. A., Extraordinary optical transmission through sub-wavelength hole arrays. Nature 1998, 391 (6668), 667-669.
\bibitem {Fang05}Fang, N.; Lee, H.; Sun, C.; Zhang, X., Sub-Diffraction-Limited Optical Imaging with a Silver Superlens. Science 2005, 308 (5721), 534-537.
\bibitem {Liedberg83}Liedberg, B.; Nylander, C.; Lunstr?m, I., Surface plasmon resonance for gas detection and biosensing. Sensors and Actuators 1983, 4, 299-304.
\bibitem {Altewischer02}Altewischer, E.; van Exter, M. P.; Woerdman, J. P., Plasmon-assisted transmission of entangled photons. Nature 2002, 418 (6895), 304-306.
\bibitem {Fasel05}Fasel, S.; Robin, F.; Moreno, E.; Erni, D.; Gisin, N.; Zbinden, H., Energy-Time Entanglement Preservation in Plasmon-Assisted Light Transmission. Phys. Rev. Lett. 2005, 94 (11), 110501.
\bibitem {Ren06}Ren, X. F.; et al., Plasmon-assisted transmission of high-dimensional orbital angular-momentum entangled state. EPL (Europhysics Letters) 2006, 76 (5), 753.
\bibitem{Lezec06} Gay, G.; Alloschery, O.; Viaris de Lesegno, B.; O/'Dwyer, C.; Weiner, J.; Lezec, H. J., The optical response of nanostructured surfaces and the composite diffracted evanescent wave model. Nat Phys 2006, 2 (4), 262-267.
\bibitem{Lopez07} Lopez-Tejeira, F.; Rodrigo, S. G.; Martin-Moreno, L.; Garcia-Vidal, F. J.; Devaux, E.; Ebbesen, T. W.; Krenn, J. R.; Radko, I. P.; Bozhevolnyi, S. I.; Gonzalez, M. U.; Weeber, J. C.; Dereux, A., Efficient unidirectional nanoslit couplers for surface plasmons. Nat Phys 2007, 3 (5), 324-328.
Dereux, Nature Phys. 3, 324 (2007).
\bibitem{Paci07} Pacifici, D.; Lezec, H. J.; Atwater, H. A., All-optical modulation by plasmonic excitation of CdSe quantum dots. Nat Photon 2007, 1 (7), 402-406.
\bibitem {Zia07}Zia, R.; Brongersma, M. L., Surface plasmon polariton analogue to Young's double-slit experiment. Nat Nano 2007, 2 (7), 426-429.
\bibitem {Ren08}Ren, X.-F.; Guo, G.-P.; Huang, Y.-F.; Wang, Z.-W.; Zhang, P.; Guo, G.-C., Interference of surface plasmon polaritons controlled by the phase of incident light. Appl. Phys. Lett. 2008, 92 (17), 171106-3.
\bibitem{Dall09}Dallapiccola, R.; Dubois, C.; Gopinath, A.; Stellacci, F.; Dal Negro, L., Near-field excitation and near-field detection of propagating
surface plasmon polaritons on Au waveguide structures,
Appl.Phys.Lett. 2009, 94, 243118.
\bibitem {Bouhelier07}Bouhelier, A.; Ignatovich, F.; Bruyant, A.;
Huang, C.; Colas des Francs, G.; Weeber, J. C.; Dereux, A.;
Wiederrecht, G. P.; Novotny, L., Surface plasmon interference
excited by tightly focused laser beams. Opt. Lett. 2007, 32 (17),
2535-2537.
\bibitem {Liu05}Liu, Z.; Steele, J. M.; Srituravanich, W.; Pikus, Y.; Sun, C.; Zhang, X., Focusing Surface Plasmons with a Plasmonic Lens. Nano Letters 2005, 5 (9), 1726-1729.
\end{thebibliography}
\end{document}